\documentclass[showpacs,aps,twocolumn,prb,floatfix,superscriptaddress,citeautoscript]{revtex4-1}
\usepackage{amsmath}
\usepackage{amssymb}
\usepackage{graphicx}

\usepackage{epsf}
\usepackage{epsfig}
\usepackage{dcolumn}
\usepackage{braket}
\usepackage{bm}
\usepackage{amsfonts}
\usepackage{color,soul}
\usepackage{times}

\newcommand{\ve}{\varepsilon}

\newcommand{\Ef}{\ve_{\text{F}}}
\newcommand{\kt}{k_{\text{B}}T}
\newcommand{\bea}{\begin{eqnarray}}
\newcommand{\eea}{\end{eqnarray}}

\newcommand{\ci}{\mathrm{i}}

\begin{document}

\title{Partial preservation of chiral symmetry and colossal magnetoresistance in adatom doped graphene}
\author{Gonzalo Usaj}
\affiliation{Centro At{\'{o}}mico Bariloche and Instituto Balseiro, CNEA, 8400 Bariloche, Argentina}
\affiliation{Consejo Nacional de Investigaciones Cient\'{\i}ficas y T\'ecnicas (CONICET), Argentina}
\author{Pablo S. Cornaglia}
\affiliation{Centro At{\'{o}}mico Bariloche and Instituto Balseiro, CNEA, 8400 Bariloche, Argentina}
\affiliation{Consejo Nacional de Investigaciones Cient\'{\i}ficas y T\'ecnicas (CONICET), Argentina}
\author{C. A. Balseiro}
\affiliation{Centro At{\'{o}}mico Bariloche and Instituto Balseiro, CNEA, 8400 Bariloche, Argentina}
\affiliation{Consejo Nacional de Investigaciones Cient\'{\i}ficas y T\'ecnicas (CONICET), Argentina}
\date{submitted 28 November 2013, accepted 24 January 2014}

\begin{abstract}
We analyze the electronic properties of adatom doped graphene in the low impurity concentration regime. 
We focus on the Anderson localized regime and calculate the localization length ($\xi$) as a function of the electron doping and an external magnetic field. 
The impurity states hybridize with carbon's $p_z$ states and form a partially filled band close to the Dirac point. 
Near the impurity band center, the chiral symmetry of the system's effective Hamiltonian is partially preserved which leads to a large enhancement of  $\xi$. The sensitivity of transport properties, namely Mott's variable range hopping scale $T_0$,  to an external magnetic field perpendicular to the graphene sheet leads to a colossal magnetoresistance effect, as observed in recent experiments.
\end{abstract}

\maketitle

The peculiar electronic structure of graphene, with chiral quasiparticles behaving as
massless Dirac fermions, gives rise to a number of remarkable and counterintuitive phenomena \cite{CastroNeto-review,*DasSarma2011,*Beenakker2008} that  manifest both in  pristine  and  disordered graphene \cite{Evers2008}. 

In disordered systems, electron localization depends on dimensionality and on the nature of disorder \cite{Lee1985,Evers2008}. Due to the particular symmetries of graphene and the way different types of disorder break these symmetries, the problem of electron localization requires revisiting some of the basic and conceptual issues \cite{Aleiner2006,Ostrovsky2006,*Mirlin2010,*Ostrovsky2007}.
Much has been done during the last years in this direction and it is now clear that in the absence of short range disorder Dirac fermions elude Anderson localization as in that case the system belongs to the symplectic universality class.
Short range disorder due to defects at the atomic scale generates inter-valley mixing and breaks the symplectic symmetry. Within this scenario, the symmetry of functionalized graphene belongs to the orthogonal universality class and, like in other more conventional two-dimensional (2D) systems, Anderson localization might occur. However, properties at zero energy, the Dirac point (DP), are peculiar with adatoms and vacancies leading to different behaviour \cite{Ostrovsky2006,*Mirlin2010,*Ostrovsky2007,Konig2012,bipartita}.

It is well known that the localization properties of 2D materials can be studied by applying a perpendicular (out of plane) magnetic field that suppress the quantum interference effects responsible for the electron localization \cite{Pichard1990,*Bouchaud1991,*Lerner1995}. The magnetic field  can also introduce orbital effects for large fields \cite{Raikh1992,Li1989,Jiang1992}.
In the case of graphene, one might then expect an anomalous behaviour of the localization  \cite{Ortmann2011,Ortmann2013,Cresti2013,Gattenloehner2013} or the transport properties \cite{Hong2011} since  the Landau levels (LLs) present an unusual spectrum with the zeroth LL ($0$-LL) pinned to the DP and a large energy splitting between LLs. 

In practice short range disorder can be controlled by chemical functionalization, hydrogenation \cite{Matis2012,Guillemette2013} and fluorination \cite{Hong2011} being among the most studied cases although adsorption of transition metal atoms, oxygen and molecules have also been considered \cite{Wehling2010,Roche2012}. 
Most of these defects, either adatoms or vacancies, generate resonant states close to the DP \cite{Pereira2006} and, with the appropriate concentration, may lead to strong localization regimes at low energies \cite{Yuan2012,Cresti2013}.

Here, we analyse the problem of electron localization in graphene with diluted impurities, both in the absence and in the presence of a magnetic field, using a model suitable for the description of adatoms, which are represented by a single level of energy $\varepsilon_0$ hybridized to the carbon's $p_z$ states \cite{Pereira2006,Wehling2010b,Sofo2011}. Our results show that: (\textit{i}) the localization length presents a maximum near, but not at, the DP which is reminiscent of the anomalous behaviour expected at the DP for $\varepsilon_0=0$ impurities \cite{Ostrovsky2006,*Mirlin2010,*Ostrovsky2007,Konig2012,Yuan2012,Cresti2013}; (\textit{ii}) the magnetic field leads to a large increase of the localization length in a magnitude that is consistent with the magnetoresistance found in fluorinated graphene \cite{Hong2011}.

The Hamiltonian of the system
is given by $H=H_{0}+H_\mathrm{F}+H_\mathrm{hyb}$. The  first term describes the graphene sheet
$H_{0}=-\sum_{\langle i,j\rangle}(t_{ij}\,c_{i}^{\dag}c_{j}^{}+h.c.)$
where $c_{i}^{\dag }$ creates an electron on site $i$ of the honeycomb lattice---we here neglect the Zeeman coupling and drop the spin index in what follows. The orbital effect is included through the
Peierls substitution for the hoping matrix element $t_{ij}=te^{-\ci\varphi
_{ij}}$with $t=2.8$ eV and $\varphi_{ij}$ a gauge dependent phase 
$
\varphi_{ij}=\frac{2\pi }{\phi _{0}}\int_{\bm{R}_{j}}^{\bm{R}_{i}}\bm{A}\cdot d%
\bm{\ell}
$
where $\phi _{0}$ is the flux quantum, $\bm{A}$ the vector potential and
$\bm{R}_{i}$ is the coordinate of site $i$. 
We consider impurities which are adsorbed on top of carbon atoms and are described by
$H_\mathrm{F}=\sum_{l}^{\prime}\varepsilon_{0}\,f_{l}^{\dag
}f_{l}^{}$
where $f_{l}^{\dag }$ creates an electron  on the impurity orbital of the atom at site $l$, and the primed sum runs over the indices of carbon atoms having an impurity on top.
The last term of the Hamiltonian describes the hybridization of the impurity
and the graphene orbitals
$H_{\mathrm{hyb}}=V\sum_{l}^{\prime}f_{l}^{\dag }c_{l}^{}+h.c.$
We consider systems with a low concentration $n_{i}$ of impurities,
typically $n_{i}\lesssim10^{-3}$, and  take $V=2t$.

\begin{figure}
\includegraphics[width=.95\columnwidth]{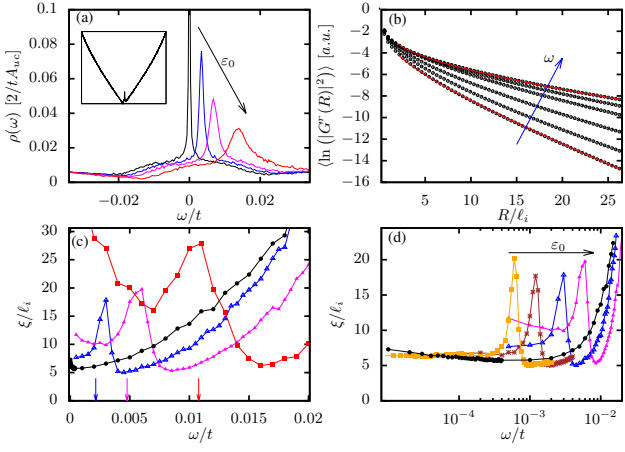}
\caption{(color online) (a) Average local density of states near the DP for increasing values of the impurity level energy, $\varepsilon_0/t=0,0.025, 0.05,0.1$ (the inset shows a zoom out for $\varepsilon_0=0.05$). A peak in the density of states forms near the renormalized energy $\bar{\varepsilon}_0$. The impurity concentration is $n_i=1/1800$ (b) Spatial dependence of $\langle\ln{|G^r_{ij}(\omega )|^2}\rangle_\mathrm{avg}$ inside the impurity band for $\varepsilon_0=0$. The solid lines are fittings to Eq. (\ref{lnG}) (taking $\alpha=1$) for two cases, $\omega=0.002t$ and $0.012t$. The localization length extracted from all these curves is shown in (c) for $\varepsilon_0/t=0\, (\bullet),0.025\,(\triangle), 0.05\, (\blacktriangle),0.1\,(\blacksquare)$. Lines are guides to the eye. The arrows shown the position of $\bar{\varepsilon}_0$. (d) Same as before but including the cases of $\varepsilon_0/t=0.005\, (\times),0.01\,(\ast)$ and using a log-scale for $\omega$. }
\label{dos}
\end{figure}

We define the impurity, $\mathcal{G}_{ij}^{r}=\langle\langle f_{i}^{},f_{j}^{\dag }\rangle\rangle $, and the gra\-phe\-ne, $G_{ij}^{r}=\langle\langle c_{i}^{},c_{j}^{\dag }\rangle\rangle$, retarded propagators. 
The average local density of states (LDOS) at a carbon site is then given by
\begin{equation}
\rho^{c}_{}(\omega)=-\frac{1}{\pi}\langle\mathrm{Im}G_{ii}^{r}\rangle_\mathrm{avg}\,,
\end{equation}
where $\langle\dots\rangle_\mathrm{avg}$ indicates the configurational average over the impurities. A similar expression gives the impurity LDOS $\rho^{f}(\omega)$. The total average DOS per atom is given by $\rho(\omega)=[\rho^{c}(\omega)+n_{i}\,\rho^{f}(\omega)]/(1+n_{i})$.
To calculate $\rho(\omega)$ we use the Chebyshev polynomials method which is very well suited to deal with realistic impurity concentrations \cite{Weisse2006,Covaci2010,Yuan2010}. 
Figure \ref{dos}(a) shows the LDOS in the absence of a magnetic field, $\mathcal{B}=0$, for a system of $\sim3\times10^7$ C atoms with $n_i\sim5\times 10^{-4}$ and for different values of $\varepsilon_0$. A new peak emerges in the LDOS \cite{Pereira2006,Peres2006,Yuan2010,Wehling2010b} which is located near the renormalized energy $\bar{\varepsilon}_0=\varepsilon_0+\Re\Sigma(\bar{\varepsilon}_0)$,  where $\Sigma(\omega)$ is the single impurity self-energy \cite{Uchoa2008,*CastroNeto2009a,Sofo2012}. For the impurity parameters used in this work, the renormalized energy is an order of magnitude lower than the bare energy ($\bar{\varepsilon}_0\ll\varepsilon_0$).
For $\varepsilon_0=0$, the peak lies at the DP ($\omega=0$) as the electron-hole symmetry is preserved. 
In what follows we will refer to the states associated to the peak in the LDOS as the `impurity band', although it is important to emphasize that the corresponding eigenstates involve both impurity and C atoms, being in general a superposition of many single impurity states \cite{Pereira2006,Sofo2011,Sofo2012}
As we now show, these states are localized and, consequently, when the chemical potential lies  within this band the system is insulating. Assuming that each adatom has originally a single electron in the relevant atomic level, the chemical potential for ungated graphene satisfies this condition for the impurity parameters considered below. 

To estimate the localization length
$\xi (\omega )$ we evaluate the two-point correlation function $|G_{ij}^r(\omega)|^2$, 
 where $G_{ij}^{r}$ is the retarded propagator 
from site $i$  to site $j$. In the localized regime this quantity decreases exponentially when the distance $R_{ij}$ between sites increases. For large $R_{ij}$ ($R_{ij}\gtrsim\xi$), the configurational average of its logarithm is well described by the following expression \cite{Li1989}
\begin{equation}
\label{lnG}
\langle\ln{|G^r_{ij}(\omega )|^2}\rangle_\mathrm{avg}=\beta-2R_{ij}/\xi (\omega )-\alpha \ln{R_{ij}}\,.
\end{equation}
We define the impurity propagator matrix $\bm{\mathcal{G}}$ with matrix elements $\mathcal{G}^r_{ij}(\omega)$. In this notation, the  equation of motion (or Dyson equation) reads
\begin{equation}
\left[(\omega+\ci0^+-\varepsilon _{0})\bm{I}-V^{2}\tilde{\bm{g}}\right]\bm{\mathcal{G}}=\bm{I}\,,
\end{equation}
where $\bm{I}$ is the unit matrix and $\tilde{\bm{g}}$ is a matrix whose elements are
the propagators of pristine graphene, $g_{ij}^r$, between impurity sites---then $\tilde{t}_{ij}=V^2g_{ij}^r$ represents an effective coupling between impurities. For $n_i\ll1$, the average distance between
impurities is much larger than the lattice parameter and the propagators $g_{ij}^r$ can be approximated by the corresponding analytical expressions in the continuum limit \cite{Basko2008,*Rusin2011}.
We define a cluster of $N$ impurities, typically $N\sim\!7000$, located at random positions inside a disc of radius $\sim (a_0/2)\sqrt{N/n_i\pi}$, $a_0$ is the C-C distance,  and invert the matrix $(\omega +\ci0^+-\varepsilon _{0})\bm{I}-V^{2}\tilde{\bm{g}}$ to obtain $\bm{\mathcal{G}}$. 
In terms of these quantities, $\bm{G}$---in matrix notation, with elements $G_{ij}^r(\omega)$---is given by
\begin{equation}
\bm{G}=\bm{g+}V^{2}\bm{g}\bm{\mathcal{G}}\bm{g}\,.
\label{G}
\end{equation}
Note that $\bm{G}$ is not restricted to the impurity sites, so Eq. (\ref{G}) involves the pristine matrix propagator $\bm{g}$ that connects arbitrary C sites. 
We use a realistic concentration $n_i=1/1800$ \cite{Hong2011}, which leads to an average inter-impurity distance $\ell_i\sim50a_0$. 
Figure \ref{dos}(b) shows $\langle\ln|G^r_{ij}(\omega )|^2\rangle_\mathrm{avg}$ vs $R_{ij}$ for $\varepsilon_0=0$ and different values of $\omega$. The solid lines are fits using Eq. (\ref{lnG}) (for fixed $\alpha=1$) from where the localization length is obtained. As expected, identical results for $\xi (\omega )$ are obtained using 
$\mathcal{G}^r_{ij}$.

Figure \ref{dos}(c) shows $\xi(\omega)$ for different values of the impurity energy ($\varepsilon_0$).
 In the special particle-hole symmetric case ($\varepsilon_0=0$, circles) the localization length  increases away from the DP roughly as $\omega^2$. For the very low impurity concentration considered
 it is necessary to reach energies smaller than $\sim\!10^{-3} t$ to observe the expected increase of $\xi(\omega)$ near the DP due to the chiral symmetry of the problem \cite{Ostrovsky2006,*Mirlin2010,*Ostrovsky2007,Yuan2012,Konig2012,Cresti2013}. In fact, Fig.~\ref{dos}(d) suggest that $\xi(\omega)$ increases logarithmically as $\omega\!\rightarrow\!0$ \cite{Marko2010,*Schweitzer2012}. Near the edge of the impurity band ($\omega\!\sim\! 0.02t$, see Fig.~\ref{dos}(a)), and above, we do not obtain a clear exponential behaviour suggesting that at these energies a weak localization regime sets in as observed in Ref. [\onlinecite{Hong2011}].

For $\varepsilon_0\neq0$, we find that strong localization only exist inside the impurity band---so again, outside it (i.e. for an empty or a completely filled impurity band),
only weak localization effects are expected. Our results show that $\xi (\omega )$ presents a strong non-monotonic behaviour inside the impurity band: it shows a local maximum
close to $\bar{\varepsilon}_0$ (indicated by an arrow in Fig. \ref{dos}(c)) and a minimum slightly above the energy where $\rho(\omega)$ has its maximum---notice that the latter occurs for $\omega>\bar{\varepsilon}_0$.
We interpret this effect as governed by the same physics which leads to a reduced localization in the presence of chiral symmetry (as occurs in the $\varepsilon_0=0$ case) \cite{Ostrovsky2006,Konig2012}. This is so because  the effective Hamiltonian defined by $[\bm{\mathcal{G}}(\omega\sim\bar{\varepsilon}_0)]^{-1}$ has the chiral symmetry partially preserved.
To see this it is important to notice that $|\tilde{t}_{AB}|\gg|\tilde{t}_{AA}|$ since $|g^r_{AB}(R\sim\ell_i)|\gg |g^r_{AA}(R\sim\ell_i)|$ at low energies which leads to an off-diagonal block structure of  $[\bm{\mathcal{G}}(\omega\sim\bar{\varepsilon}_0)]^{-1}$ in the `$A$-$B$' basis for the impurities. Fi\-gu\-re \ref{dos}(d) shows that the peak of $\xi(\omega)$ continuously evolves towards the DP as $\varepsilon_0$ is reduced, supporting this view. 

\begin{figure}[t]
\includegraphics[width=0.8\columnwidth]{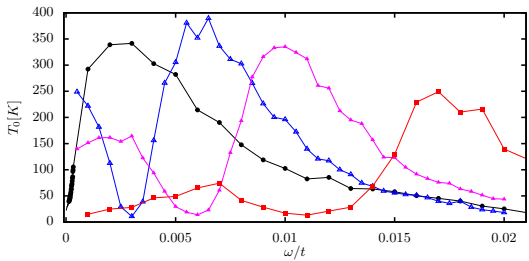}
\caption{(color online) Activation temperature $T_0$ as a function of the Fermi energy $\Ef$ for $\mathcal{B}=0$ and the same parameters as in Fig. \ref{dos}(c).}
\label{T0B0}
\end{figure}

In the strong localization regime, the resistance $\mathcal{R}(T)$ is expected to show the Mott´s variable range hopping behaviour of a $2$D system, \textit{i.e.} 
$\mathcal{R}(T)\propto \exp[(T_{0}/T)^{\frac{1}{3}}]$, where $T_0$ is a cha\-rac\-teristic activation temperature given by  
\begin{equation}
T_{0}=\frac{\gamma}{\kt\rho(\Ef)\xi ^{2}(\Ef)}\,,
\end{equation}
with $\gamma$  a numerical constant from percolation theory ($\gamma\approx 14$) and $\Ef$  the Fermi energy. 
We emphasize that the states involved in the variable range hopping processes correspond to the localized eigenstates of the full Hamiltonian described above and not to the power-law decaying single impurity states as assumed in [\onlinecite{Liang2012}]. 
Fig. \ref{T0B0} shows $T_{0}$ as a function of energy for different values of $\varepsilon_0$. In all cases the maximum value of $T_0$ is slightly shifted from the minimum value of $\xi$. Notice that, with this parameters, $T_{0}$ attains a maximum value of $\sim 300K$ which is close to the one observed in Ref. [\onlinecite{Hong2011}]. 

Let us now discuss the effect of a perpendicular magnetic field $\mathcal{B}$. In this case, the resistance is expected first to decrease, as a result of an increase of the localization length \cite{Pichard1990,Bouchaud1991,Lerner1995}, and show a crossover to a different regime when  $\mathcal{B}$ is large enough so that the magnetic length $\ell_B=\sqrt{\hbar c/e\mathcal{B}}$ gets of the order of $\ell_i$ and the shrinking of the wave function precludes the coupling between impurities, $|g^r_{ij}(R_{ij}>\ell_B)|\sim0$. 
We analyze the regime $\ell_B\gtrsim\ell_i$ using the same methods as above to calculate both $\rho$ and $\xi$ in the presence of $\mathcal{B}\ne0$.

\begin{figure}[t]
\includegraphics[width=.95\columnwidth]{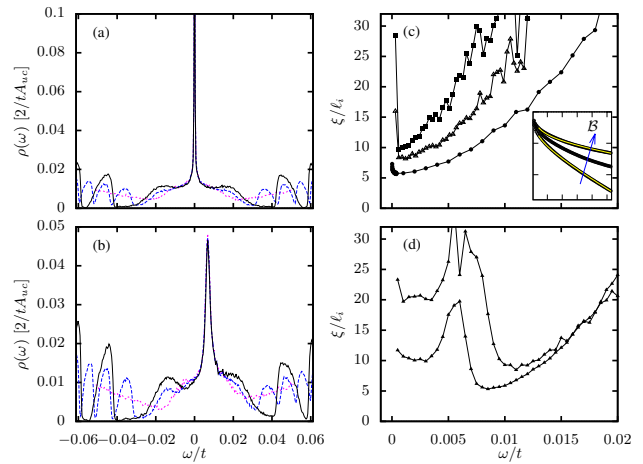}
\caption{(color online) (a),(b) Average density of states for $\varepsilon_0=0$ [(a)] and $\varepsilon_0=0.05t$ [(b)] and $\mathcal{B}=0,6$ and $12$T (dotted, dashed and solid lines, respectively). Note the splitting of the zeroth Landau level. (c) $\xi(\omega)$ for the values of $\mathcal{B}$ shown in (a); the inset  shows the spatial dependence of $\langle\ln{|G^r_{ij}|^2}\rangle_\mathrm{avg}$ for $\omega/t=5\times 10^{-3}$ and increasing values of $\mathcal{B}$. The solid lines are fittings to Eq. (\ref{lnG}) for $\mathcal{B}=0$ and $20$T. (d)  $\xi(\omega)$ for  two of the cases ($\mathcal{B}=0$ and $6$T) shown in (b).
}
\label{dosB}
\end{figure}

Figures \ref{dosB}(a) and \ref{dosB}(b) present $\rho(\omega)$ for two cases, $\varepsilon_0=0$ and $\varepsilon_0=0.05t$, respectively, and three values of the magnetic field: $\mathcal{B}=0,6$ and $12$ T. In both cases the emergence of LLs is apparent in the figure as expected. Note, however, the di\-ffe\-ren\-ce in the broadening of the LLs at the two sides of the impurity band in the case $\varepsilon_0\neq0$---this asymmetry increases with increasing $|\ve_0|$. More interestingly, the 0-LL is split by the impurities \cite{Ortmann2013,Gattenloehner2013}, manifested by the  `shoulders' that $\rho(\omega)$ develops near the edge of the impurity band with increasing va\-lues of $\mathcal{B}$. This results from the coupling of the  impurity's orbitals and the $0$-LL states located near each impurity site.

Figures \ref{dosB}(c) and \ref{dosB}(d) show the energy dependence of the localization length at different external magnetic fields for the parameters of figures \ref{dosB}(a) and \ref{dosB}(b), respectively.    
The inset to Fig. \ref{dosB}(c) shows the spatial decay of $\langle\ln{|G^r_{ij}|^2}\rangle_\mathrm{avg}$ for increasing values of $\mathcal{B}$ from where the increase of the localization length is clear \cite{Gattenloehner2013}. This increment is quantified in Fig. \ref{dosB}(c) where we show a comparison of $\xi(\omega)$ for different values of $\mathcal{B}$. The values of $\xi(\omega)$ where obtained by fitting the numerical data with Eq. (\ref{lnG})---leaving now $\alpha$ as a free parameter. Our results show that $\xi$ increases with $\mathcal{B}$ in the whole range of energies inside the impurity band.
We notice that the increase in $\xi$ can be rather dramatic, in particular close to $\bar{\ve}_0$, where $\xi$ reaches its maximum value inside the impurity band.

\begin{figure}[tb]
\includegraphics[width=.89\columnwidth]{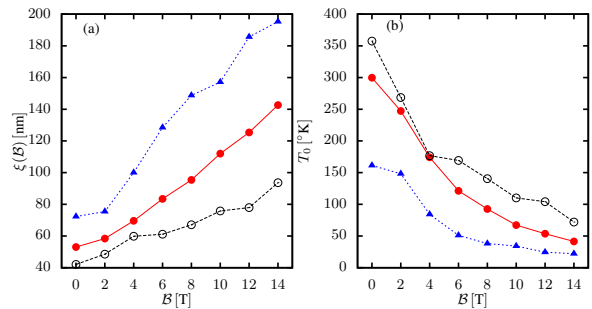}
\caption{(color online) (a) Localization length as a function of the magnetic field for the electron-hole symmetric case ($\ve_0=0$) for $\omega/t=2\times10^{-3}(\circ)$, $5\times10^{-3}(\bullet)$ and $8\times10^{-3}(\blacktriangle)$. (b) Characteristic activation temperature $T_0$ for the same energies.}
\label{T0}
\end{figure}

The increase of the localization length with magnetic field  for $\omega/t=2, 5, 8\times10^{-3}$ is shown in Fig. \ref{T0}(a) for fields up to $15$T ($\ell_B\sim\ell_i$). $\xi(\mathcal{B})$ increases by a factor $\sim3$ in this range of $\mathcal{B}$. 
As mentioned above, this increment is expected on general grounds due to the breaking of time reversal symmetry, and the consequent suppression of the interference effects that lead to localization---the magnitude of the increment on 2D systems is not universal unlike the 1D case  \cite{Pichard1990,*Bouchaud1991,*Lerner1995}.

 In the graphene case in particular, there is also a rather peculiar orbital effect that, as we numerically verified, contributes to the delocalization effect but that it is difficult to disentangle from the previous `phase factor' effect. Namely, the impurity states are always very close in energy to the $0$-LL, which is pinned to the DP. Therefore, at low impurity densities, a rather modest magnetic field is sufficient to have $\hbar\omega_c$ larger than the impurity bandwidth. In such a case, the properties of the system are mostly determinated by the pristine Green's function corresponding to the $0$-LL states, that has the particular property of not mixing different sublattices. Consequently, the network of effectively coupled impurities is changed with $\mathcal{B}$ as $\tilde{t}_{ij}$ is substantially different for sites on the same or different sublattices.

The increment of $\xi(\mathcal{B})$ leads to a decrease of $T_0$ as shown in Fig. \ref{T0}(b). It is important to point out that our results correspond to a \textit{fixed} value of $\omega$ while the experimentally relevant scenario requires to tune $\Ef$ in order to keep the electron density constant. For the parameters of Fig. \ref{T0},  this corresponds to an interpolation between the curves of, say, $\omega/t=5\times10^{-3}$ and $8\times10^{-3}$ as we increase the field form $0$ to $8$-$10$T---the fact the $\Ef$ slightly increases with $\mathcal{B}$ is related to the splitting of the $0$-LL that transfers some spectral weight from the DP to higher energies.
Once this correction is taken into account,  the change of $T_0$  with magnetic field is in quantitative agreement with the experimental data of Ref. [\onlinecite{Hong2011}]. 

In summary, we have shown that the peculiar localization properties induced by adatoms on graphene not only manifest in electron-hole symmetric systems ($\varepsilon_0=0$) but also in the general case `near' the center of the impurity band ($\omega\sim\bar{\varepsilon}_0$). In addition, we found that these properties change in the presence of a magnetic field in a manner that is in quantitative agreement with existent experimental data.
Since our model does not include any spin related effect (adatom induced magnetism or spin-orbit coupling), we conclude that the magnetoresistance data alone (in the strongly localized regime) does not provide enough evidence to support that spin-flip processes play a mayor role \cite{Hong2011,Hong2012,Kim2013} and further studies are necessary to settle this issue.

We acknowledge financial support from PICTs 2006-483, Bicentenario 2010-1060 from ANPCyT, PIP 11220080101821 from CONICET, Argentina and grant 06-C400 from UNCuyo. GU acknowledges useful discussions with T. Wehling, P. M. Ostrovsky and I. V. Gornyi.

%

\end{document}